\documentclass[twocolumn]{aastex63}

\usepackage{epstopdf}
\usepackage{multimedia}
\usepackage{epsfig,graphics,subfigure,psfrag,amsmath,amssymb,amscd}
\usepackage{url}

\begin{document}

\title{The X-ray outburst of PG 1553+113: A precession effect of two jets in the supermassive black hole binary system}

\author{Shifeng Huang}
\affiliation{Shandong Key Laboratory of Optical Astronomy and Solar-Terrestrial Environment,\\
 School of Space Science and Physics, \\
 Institute of Space Sciences, Shandong University,  \\
 Weihai, Shandong, 264209, China.
 \href{Corresponding author.}{husm@sdu.edu.cn, yinhx@sdu.edu.cn}}

\author{Hongxing Yin}
\affiliation{Shandong Key Laboratory of Optical Astronomy and Solar-Terrestrial Environment,\\
 School of Space Science and Physics, \\
 Institute of Space Sciences, Shandong University,  \\
 Weihai, Shandong, 264209, China.
 \href{Corresponding author.}{husm@sdu.edu.cn, yinhx@sdu.edu.cn}}

\author{Shaoming Hu}
\affiliation{Shandong Key Laboratory of Optical Astronomy and Solar-Terrestrial Environment,\\
 School of Space Science and Physics, \\
 Institute of Space Sciences, Shandong University,  \\
 Weihai, Shandong, 264209, China.
 \href{Corresponding author.}{husm@sdu.edu.cn, yinhx@sdu.edu.cn}}

\author{Xu Chen}
\affiliation{Shandong Key Laboratory of Optical Astronomy and Solar-Terrestrial Environment,\\
 School of Space Science and Physics, \\
 Institute of Space Sciences, Shandong University,  \\
 Weihai, Shandong, 264209, China.
 \href{Corresponding author.}{husm@sdu.edu.cn, yinhx@sdu.edu.cn}}

\author{Yunguo Jiang}
\affiliation{Shandong Key Laboratory of Optical Astronomy and Solar-Terrestrial Environment,\\
 School of Space Science and Physics, \\
 Institute of Space Sciences, Shandong University,  \\
 Weihai, Shandong, 264209, China.
 \href{Corresponding author.}{husm@sdu.edu.cn, yinhx@sdu.edu.cn}}

\author{Sofya Alexeeva}
\altaffiliation{LAMOST fellow}
\affiliation{CAS Key Laboratory of Optical Astronomy, National Astronomical Observatories, Beijing, 100101, China.
 }

\author{ Yifan Wang}
\affiliation{Shandong Key Laboratory of Optical Astronomy and Solar-Terrestrial Environment,\\
 School of Space Science and Physics, \\
 Institute of Space Sciences, Shandong University,  \\
 Weihai, Shandong, 264209, China.
 \href{Corresponding author.}{husm@sdu.edu.cn, yinhx@sdu.edu.cn}}

\begin{abstract}
Blazar PG 1553+113 is thought to be a host of supermassive black hole binary (SMBHB) system. A 2.2-year quasi-periodicity in the $\gamma$-ray light curve was detected, possibly result of jet precession. Motivated by the previous studies based on the $\gamma$-ray data, we analyzed the X-ray light curve and spectra observed during 2012--2020. The 2.2-year quasi-periodicity might be consistent with the main-flare recurrence in the X-ray light curve. When a weak rebrightening in the $\gamma$-ray was observed, a corresponding relatively strong brightening in the X-ray light curve can be identified. The ``harder-when-brighter" tendency in both X-ray main and weak flares was shown, as well as a weak ``softer-when-brighter" behavior for the quiescent state.
We explore the possibility that the variability in the X-ray band can be interpreted with
two-jet precession scenario. Using the relation between jets and accretion discs, we derive the primary black hole mass $\simeq 3.47\times 10^8M_{\sun}$ and mass of the secondary one $\simeq 1.40\times 10^8M_{\sun}$, and their mass ratio $\sim 0.41$.

\end{abstract}


\section{Introduction}
The class of active galactic nuclei (AGNs) with their jets pointed at the observer, is classified as blazars \citep{urry1995}.
At the center of blazars, supermassive black holes (SMBHs) play the role of engine triggering the jet. Blazars emit radiation over a wide range of frequencies, from radio to $\gamma$-rays.

PG 1553+113 is a blazar with redshift $z\simeq0.5$ \citep{tavani2018} and it most probably hosts a SMBHB \citep{caproni2017,cavaliere2017,sobacchi2017} resulting in the quasi-periodicity about $2.2~\text{yr}$ in the $\gamma$-ray light curves \citep{ackermann2015,yan2018,sandrinelli2018,covino2020}.
The mechanism of the quasi-periodicity in light curves of blazars is still under debates and the following scenarios were considered:
the geometric origin such as jet precession \citep{abraham1998,abraham1999,abraham2000,britzen2018}, lighthouse effect \citep{camenzind1992}, beaming effect \citep{villata1998} and helical structure of the jet \citep{zhou2018}.
However, the contribution from the disc should also be considered, such as the black hole-disc impaction \citep{lehto1996,dey2018}, which successfully predicted the 12-year optical quasi-periodic outburst in OJ 287 \citep{valtonen2008,valtonen2016,laine2020}. Moreover, the disc instability could be one of the possibilities \citep{tanaka2013} as well.

The quasi-periodicity in the PG 1553+113 light curves is probably related to a jet precession in the SMBHB system \citep{caproni2017,sobacchi2017}.
However, the $\gamma$-ray light curve shows weaker flares near the main ones that might indicate the signs of the twin jets in the system \citep{tavani2018,cavaliere2019}.

 Recently, there have been many investigations of the quasi-periodicity for PG 1553+113 mainly based on the $\gamma$-ray light curves. In this work, we focus not only on the $\gamma$-ray band, but also X-ray. The differences between X-ray and $\gamma$-ray light curves are due to prominent flares in the X-ray light curve while weaker counterparts are detected in $\gamma$-ray.
 In addition, the interval of each X-ray flare is much shorter than the 2.2 year quasi-periodicity in the $\gamma$-ray light curve. Motivated by \cite{tavani2018}, we consider the two-jet structure in this SMBHB system. The motion of a jet ejected by one of two black holes results in the quasi-periodicity of the multi-wavelength light curves. While, the jet ejected by another black hole also contributes to the variation, and the phase of the two jets leads to a shorter interval of the flares. The model was employed to study the double-peak structure in the optical outburst of OJ 287 \citep{villata1998}, and in this study, we apply this model to the case of PG 1553+113 and the results show that it is also a possible model.

This paper is organized as follows. In section 2, data reduction is introduced and the results are shown in section 3. We fit the X-ray light curve with the model and further discussion is in section 4. Conclusions are presented in section 5. The following cosmology parameters are used in this work: $H_0=70~\text{km}~\text{s}^{-1}\text{Mpc}^{-1}$, $\Omega_M=0.27${\bf,} and $\Omega_{\Lambda}=0.73$.

\section{Observation and Data Reduction}
\subsection{Fermi $\gamma$-ray data}

 The Large Area Telescope (\emph{Fermi}-LAT), the primary instrument on the Fermi Gamma-ray Space Telescope (Fermi) mission, is an imaging, wide field-of-view, high-energy $\gamma$-ray telescope, covering the energy range from below 20 MeV to more than 300 GeV \citep{atwood2009}.
We collected  $\gamma$-ray data of PG 1553+113 from the \emph{Fermi}-LAT data website\footnote{\url{https://fermi.gsfc.nasa.gov/cgi-bin/ssc/LAT/LATDataQuery.cgi}} obtained during 8 years from Jan. 1, 2012 to Jan. 1, 2020.

The data were analyzed with standard unbinned likelihood tutorials using the Fermi Science Tools \texttt{Fermitools} version 1.0.10. The photons event data were extracted by \texttt{gtselect} while selecting the source class events centered on the target from the region of interest (ROI) with the radius of $10^{\circ}$. The good time intervals were selected with the \texttt{gtmktime} procedure. A counts map of ROI was created by \texttt{gtbin} and the exposure map was generated by \texttt{gtltcube} and \texttt{gtexpmap}.

We have adopted the current Galactic diffuse emission model
(\rm{gll\_iem\_v07.fits}) and the model for the extragalactic isotropic diffuse emission iso\_P8R3\_SOURCE\_V2\_v1.txt. The source model XML file was created with Python code \rm{make4FGLxml.py}, and the diffuse source responses were created by \texttt{gtdiffrsp}. Finally, the \texttt{gtlike} procedure was run to obtain flux. The threshold of the test statistics (TS) value is set to be 10 in this work.

\subsection{Swift XRT and UVOT data}
\subsubsection{X-ray data}
The \emph{Swift} X-Ray Telescope (XRT) is one of the instruments on the Neil Gehrels \emph{Swift} observatory with a sensitive broad-band detector for X-ray from 0.3 to 10 keV \citep{burrows2005}. The data from January 1, 2012 to January 1, 2020 were retrieved from the High Energy Astrophysics Science Archive Research Center (HEASARC) website\footnote{\url{https://heasarc.gsfc.nasa.gov/db-perl/W3Browse/w3browse.pl}} and we followed standard threads\footnote{\url{https://www.swift.ac.uk/analysis/xrt/index.php}} to analyse the data from level I.

All \emph{Swift} data were reduced with the new released \texttt{HEASoft 6.26.1}. The \texttt{xrtpipeline} tool was run to reprocess the data. Centering on the target, for the windowed timing (WT) mode data, a source region file selected by a central circle with radius of 70 arcsec and a background region file was selected by an annular with an inner radius of 100 arcsec and outer radius of 150 arcsec.
The pile-up effect  was considered for the data collected in photon counting (PC) mode, and the source region was extracted by an annulus with inner and outer radius are 10 and 75 arcsec, respectively. The background region for PC mode was selected with the same method as the case of the WT. The corrected light curves and spectra were generated by \texttt{xrtproducts} with level II data from the PC mode and WT mode, respectively.

The source spectra were grouped by \texttt{grppha} with minimum 20 photons per bin for both PC and WT mode spectra. Besides, in the reduction of X-ray spectra, the redistribution matrix file swxwt0to2s6\_20131212v015.rmf for WT mode and swxpc0to12s6\_20130101v014.rmf for PC mode were used and the ancillary response files were created by \texttt{xrtmkarf}.

\subsubsection{UV-optical data}
UV/Optical Telescope (UVOT), covering wavelength range 170-600 nm, 7 filters (\textsl{v}, \textsl{b}, \textsl{u}, \textsl{uvw1}, \textsl{uvm2}, \textsl{uvw2} and white), with field of view $17^{\prime} \times 17^{\prime}$, is one of instruments on the \emph{Swift} observatory \citep{roming2005}. Following the recommended threads\footnote{\url{https://www.swift.ac.uk/analysis/uvot/index.php}}, \texttt{uvotsource} in \texttt{HEASoft 6.26.1} was run with a source circular region radius of 10 arcsec for \textsl{v}, \textsl{b}, \textsl{u} bands, but for \textsl{uvw1}, \textsl{uvw2} and \textsl{uvm2}, we used a source region file with radius of 20 arcsec. The background region files were extracted in a region without any source nearing the source with the radius of 50 arcsec. We adopted the extinction $E(B-V)=0.054$ mag, and
the results of \cite{raiteri2015} were applied here with 0.173, 0.229, 0.275, 0.394, 0.457, and 0.481 mag corresponding to the \textsl{v}, \textsl{b}, \textsl{u}, \textsl{uvw1}, \textsl{uvm2} and \textsl{uvw2} bands.

\subsection{XMM-Newton data}

\emph{XMM-Newton} is an X-ray observatory, and its main instrument is the European Photon Imaging Camera (EPIC), consisting of two MOS detectors and a pn camera which operate in the 0.2--12.0 keV energy range.
Totally, 24 sets of observations of PG 1553+113 obtained from 2001 to 2020 were employed and analysed with the software Science Analysis System (SAS version 18.0). We followed the recommended data analysis threads by XMM-Newton Science Operation Centre, \texttt{cifbuild} and \texttt{odfingest}  procedures for the preparation, and then \texttt{xmmextractor} for the extractions of the light curves and spectra. The \texttt{evselect} was run when we extracted the light curves on 0.3--2.0 keV and 2.0--10.0 keV.

\subsection{Radio data}
The 40 m telescope at the Owens Valley Radio Observatory (OVRO), is monitoring more than 1800 blazars \citep{richards2011}. We obtain the 15 GHz data covering from 2012 January to 2020 January from the public data archived website\footnote{\url{http://www.astro.caltech.edu/ovroblazars/index.php?page=home}}.

\section{Multi-wavelength light curves}
\begin{figure}
  \centering
  \includegraphics[width=0.5\textwidth]{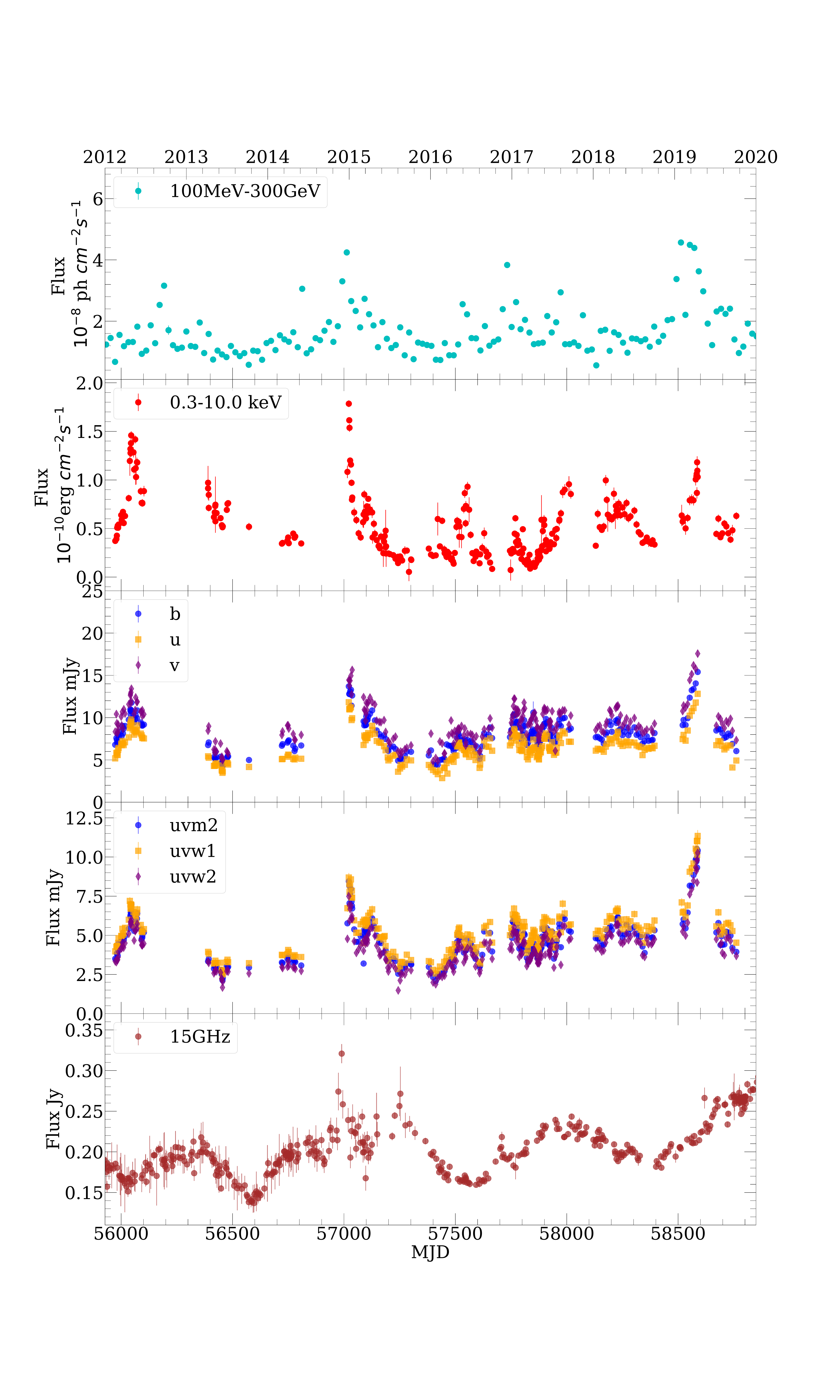}
  \caption{Multiwavelength light curves of PG 1553+113 covering $\gamma$-ray, X-ray, UV, optical and radio bands with epochs from 2012 to 2020. The $\gamma$-ray light curve is shown with a 20 day binning and the data of X-ray and optical/UV are rebinned to one data point for each observation. }\label{fig:mwlc}
\end{figure}

The multi-wavelength light curves during 2012 January -- 2020 January of PG 1553+113, ranging from $\gamma$-ray to the radio, are shown in Figure \ref{fig:mwlc}. The light curve of $\gamma$-ray exhibits the  quasi-periodic variations which were reported by \cite{ackermann2015}. Some weak flares can be seen near the main flares on the $\gamma$-ray light curve. Similar situations can also be seen in the X-ray and optical/UV bands.

When the main flare in the $\gamma$-ray light curves was observed, the prominent outbursts in both  X-ray and optical/UV bands could be seen, and, in addition, similar to the case of the $\gamma$-ray, some distinct flares in X-ray and optical/UV bands were observed when the weak flares in $\gamma$-ray were recorded.

In the radio light curve, a quasi-periodic variation can also be seen, however, weak flares are not observed clearly.

\section{Discussion}
The quasi-periodic variation in the $\gamma$-ray light curve of PG 1553+113 was explained by the jet precession in the SMBHB \citep{caproni2017}. One-jet model is appropriate for main flares, while the existence of the weak peaks on the light curve, possibly requires a two-jet model to explain this phenomenon (see  e.g., \citealt{tavani2018}, \citealt{cavaliere2019}).

We attempted to fit the X-ray light curves with the single jet precession, but unfortunately, the obtained period is much shorter than that in the case of $\gamma$-ray when the same model was used for fitting. Moreover, similar behavior of the main and weak flares in the correlation between flux and spectral index may be the signature of two jets. Based on this possibility, in order to explore the origin of the X-ray outbursts, we propose the two-jet model for this case.

\subsection{The two-jet precession model}
In this model, a SMBHB system is considered and each black hole carries  its own jet and the movement of the black holes on the orbit causes the quasi-periodic variation on the light curves \citep{villata1998,tavani2018}.

Motivated by the work of \cite{villata1998}, we obtain a function of the angle between the line of sight and the jet direction
\begin{eqnarray}
\cos\theta_{obs,1}&=&\cos\phi_1 \cos\alpha_1+\sin\phi_1 \sin\alpha_1 \cos\omega t, \\
\cos\theta_{obs,2}&=&\cos\phi_2 \cos\alpha_2+\sin\phi_2 \sin\alpha_2 \cos(\omega t+\psi),
\end{eqnarray}
where the subscripts 1 and 2 correspond to the first and the second jet, respectively, and $\theta_{obs,1}$, $\theta_{obs,2}$ are the angles between the line of sight and the jet direction, $\phi_1$ and $\phi_2$ are the angles between the line of sight and the orbital angular momentum/spin of the individual black hole (assuming that the spin parallel to the orbital angular momentum), $\alpha_1$ and $\alpha_2$ are the angles between the orbital angular momentum and the jet direction, $\omega$ is the orbital angular velocity and $\psi$ is the phase position.

The flux of the jet can be expressed as $F(t)=\delta^3(t) F_0$, where $F_0$ is
the flux in the rest frame, and $\delta(t)$ is the Doppler factor which can be written as $\delta(t)=\frac{1}{\gamma(1-\beta\cos\theta_{obs})}$, where $\beta$ is the bulk velocity and $\gamma=\left(1-\beta^2\right)^{-\frac{1}{2}}$ is the Lorentz factor. In
addition, before fitting, we transform the flux into luminosity through $L=4\pi D^2 F$, where $D$ is the luminosity distance. In our model, the observed luminosity can be obtained as
\begin{eqnarray}
L_{tot}=\delta_1^3(t)L_1+\delta_2^3(t)L_2+L_{b},
\label{eq:flux}
\end{eqnarray}
where $L_1$ and $L_2$ are the luminosity of jet 1 and jet 2, the subscript 1, 2 of $\delta$ refers to the Doppler factor of the two jets, and $L_b$ is the contribution from other sources such as the disc, corona, etc. For simplification, the precession of the disc, gravitational redshift, and intrinsic variety of accretion flow is not considered here, therefore, for convenience, we assume that $L_b$ is not a time-dependent physical quantity in our model.

\subsection{Fitting results}
\begin{figure}
  \centering
  \includegraphics[width=0.5\textwidth]{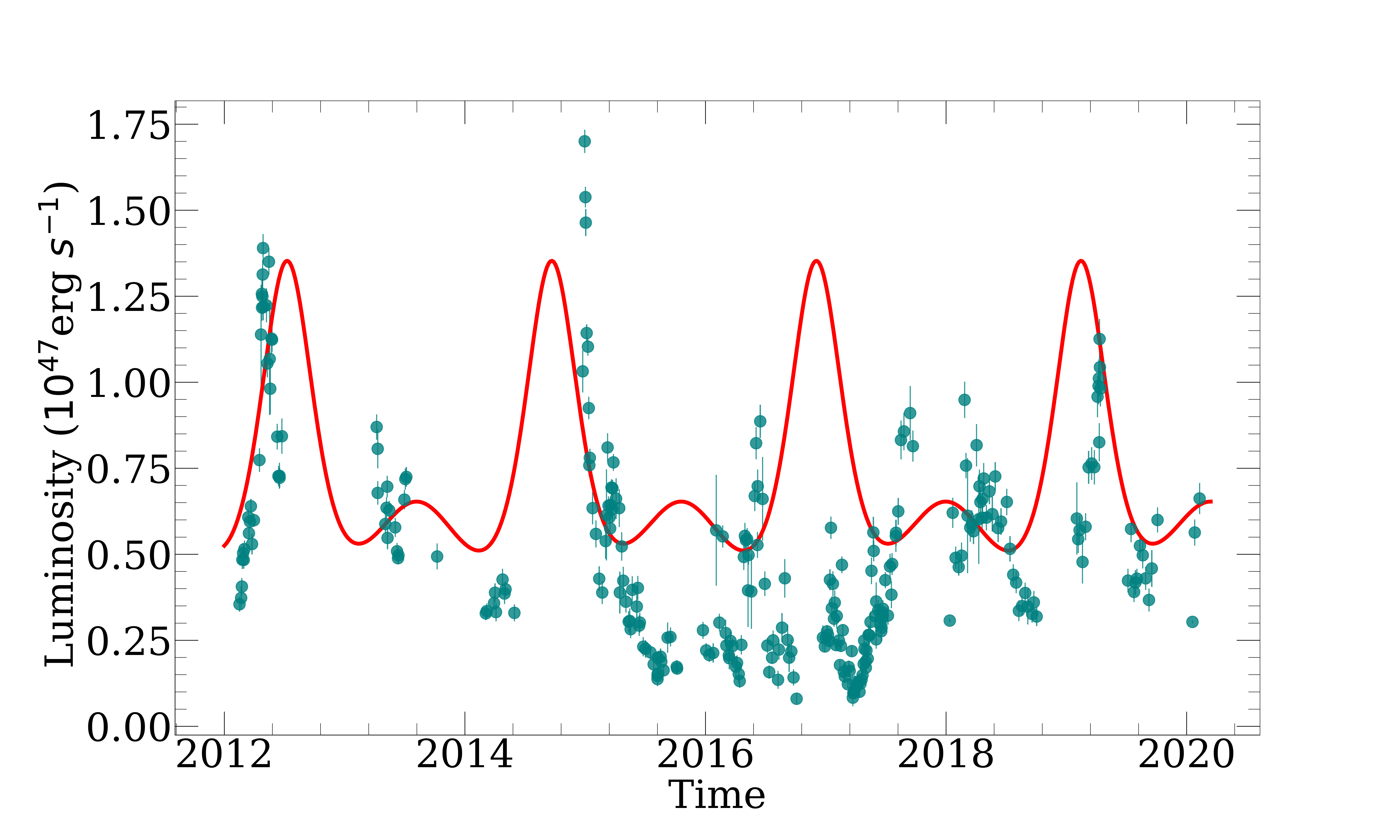}
  \caption{Fitting of the X-ray (0.3 -- 10.0 keV) light curve (dots with error bars) by the two-jet model (solid curves).  }\label{fig:x_fit}
\end{figure}

In the fitting, we only fixed the orbital angular velocities ($\omega=\frac{2\pi}{P}$) of both black holes. To be consistent with the case of $\gamma$-ray, the period was frozen as $P=2.2~\text{yr}$. Additionally, $L_b$ was frozen as $8.0 \times 10^{45} \text{erg}~\text{s}^{-1}$ which is the minimum value in the light curve. Meanwhile, with assumption $\phi_1=\phi_2$, other parameters in the model were freed. However, \cite{villata1998} assumed the two jets share the same values of the parameters when they tried to explain the quasi-periodic double peak structure in the optical light curves of OJ 287.  In our work, we do not set the parameters of the two jets (except for the orbital angular velocities and $\phi_1=\phi_2$) to be equal. The Markov Chain Monte Carlo (MCMC) \texttt{emcee} package \citep{foreman2013} was used in the fitting and 500 walkers and 10000 steps were run\footnote{The chain was accessed by using the \emph{EnsembleSampler.get\_chain} in python package \emph{emcee}. Within this method,  we obtained the parameter values as a function of the step numbers in the chain. When the parameters do not change much, the chain can be considered stable. }. We adopt the uniform prior as $0.0 < \phi_1 < 10^{\circ}$, $0.0 < \alpha_1, \alpha_2 < 10^{\circ}$, $0.0 < \psi < 2\pi$, $0.0 < \gamma_1, \gamma_2 < 50.0$, $0.0 < L_1, L_2 < 5.0\times 10^{44}\text{erg}~\text{s}^{-1}$. The fitted result is shown in Figure \ref{fig:x_fit}. More details for the fitted parameters in our model are shown in Table~\ref{tab:paremeters}.

The fitting is unsatisfactory, with a reduced chi-square of 348.41. However, we did not expect a perfect result based on the X-ray light curve, mainly because of the simple model and the lack of continuous observations for this object. Nevertheless, we had tried to free the period parameter in the fitting, but in the results, the fitted period of 2.8 years is inconsistent with the case of $\gamma$-ray \citep{ackermann2015}. \cite{tavani2018} showed the consistency between the R-band and X-ray light curves in the long-term observations. Additionally, \cite{sandrinelli2018} and \cite{tavani2018} reported a similar behavior in the long-term $\gamma$-ray and the R-band light curve. Therefore, we can assume that the $\gamma$-ray and X-ray light curves are correlated to some extent. Hence, in fitting, it is reasonable to fix the periodicity to 2.2 years from continuously monitoring $\gamma$-ray data \citep{caproni2017,sandrinelli2018,cavaliere2019}.

As mentioned in the last subsection, we only consider the geometric effect in the system in this work. The profile of the X-ray light curve of PG 1553+113 is fairly complicated, while the fitted model in this work is rather simple and it can not describe all the data points in the X-ray. We have tried to improve the model to fit the X-ray light curve and some geometric effects were considered in the test. Although the statistical fitting result is slightly improved, the data sampling still plays an important role. The sparse sample can be fitted well by such complex models may not indicate the true nature of the system. To further investigate the detail of the variability, more high cadence observations are needed. The origin of the X-ray is very complex and there exist some other mechanisms that would modulate the variability. Based on the assumption that the quasi-periodic variation in the X-ray light curve is origin from the emission of the jets, we only explore the effect of the two jets' precession. However, according to \cite{sobacchi2017}, the Lense-Thirring precession would modulate the oscillation of angle with the time scale from days to years and this may affect on the X-ray variability. Additionally, we did not consider the intrinsic variation of the jets, so our model can not mirror the activity in the jets. For these reasons, the predicted light curve may not fit the observed data very well.

Although the data points in the light curve can not be described well by the two-jet precession scenario, the main and weak flares can be explained well. In order to predict variability accurately, more aspects need to be taken into account. However, in this work, we mainly focus on the effect caused by the precession of two jets in the SMBHB system.

\begin{deluxetable*}{cccc}
\tablenum{1}
\tablecaption{The best-fit results for the parameters through two-jet model. \label{tab:paremeters}}
\tablewidth{0pt}
\tabletypesize{\scriptsize}
\tablehead{
 \colhead{Parameter} & \colhead{Value} & \colhead{Parameter} & \colhead{Value}
}
\decimalcolnumbers
\startdata
   $\phi_1$ & 4.35 degree &  $t_0$ & 2011.40\\
   $\alpha_1$ &  1.66 degree & $\alpha_2$ & 2.64 degree\\
   $\gamma_1$ & 7.67 & $\gamma_2$ & 11.96 \\
   $L_1$ & $0.21\times 10^{44}\text{erg}~\text{s}^{-1}$ & $L_2$ & $0.12\times 10^{44}\text{erg}~\text{s}^{-1}$ \\
   $\psi$ & 176.59 degree & &
\enddata
\tablecomments{We obtain the fitting results with the two-jet model through \texttt{emcee}. The subscripts 1 and 2 correspond to the parameter of the jet 1 and jet 2. $t_0$ is the reference time. }
\end{deluxetable*}

\subsection{The mass estimation of each black hole}
The mass of each black hole was not directly estimated in the previous investigations.
We can only constrain the orders of magnitude of the two black holes masses through the fitting results. Following the relation between the power of jets and the luminosity of the discs, the mass of each black hole can be estimated. According to \cite{ghisellini2014},
\begin{equation}
\log P_{rad}=0.98\log L_{disc}+0.639,
\label{eq:l_dis}
\end{equation}
where $P_{rad}$ is the radiative power of  the jet, and $L_{disc}$ is the disc luminosity. Assuming that X-ray originated from synchrotron emission, the radiative power of the jet can be written as
\begin{equation}
P_{rad}=\frac{32}{5}L_{bol,jet}\frac{\gamma^4}{\delta^6},
\label{eq:l_bol}
\end{equation}
where $L_{bol,jet}$ is the bolometric luminosity of the jet which can be estimated following the method of \cite{runnoe2012}.

Substituting the fitting results from Table \ref{tab:paremeters} into Equations (\ref{eq:l_bol}) and (\ref{eq:l_dis}), and assuming the accretion rate as 0.1 $\dot{M}_{\text{edd}}$ \citep[following][]{ackermann2015}, where $\dot{M}_{\text{edd}}$ is the Eddington accretion rate, we obtain the masses of the black holes launching jet 1 and jet 2 are $M_1\simeq 1.40\times 10^8M_{\sun}$ and $M_2\simeq 3.47\times 10^8M_{\sun}$, respectively. The inferred mass ratio is $q=\frac{M_1}{M_2}\simeq 0.41$ that is in agreement with \cite{tavani2018}. The total mass of the SMBHB system is consistent with the results of \cite{ackermann2015} and \cite{cavaliere2019}.

\subsection{The X-ray properties evolution}
\begin{figure}
  \centering
  \includegraphics[width=0.5\textwidth]{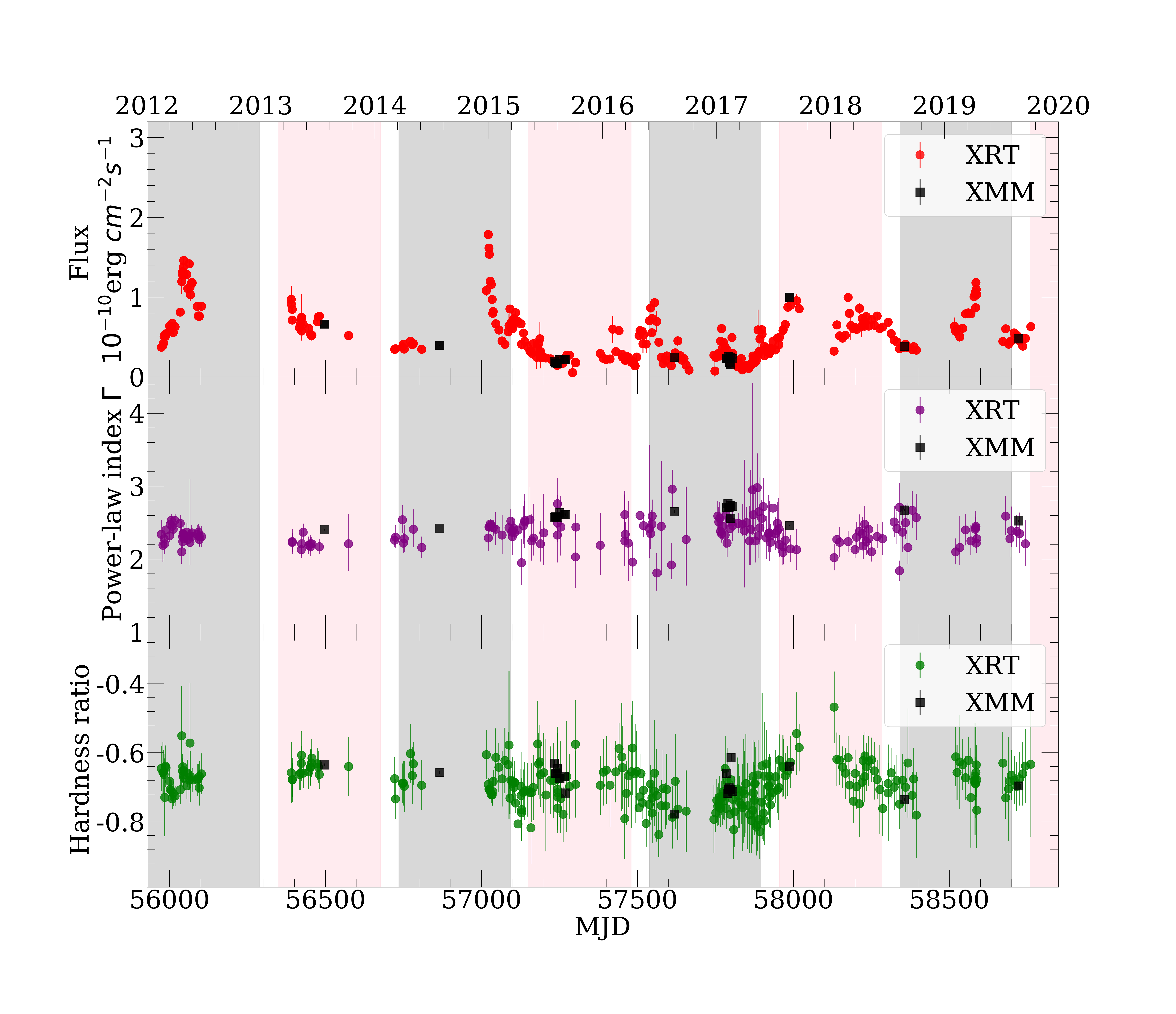}
  \caption{The evolution of X-ray (0.3--10.0 keV) flux, power-law index and hardness ratio in PG 1553+113 are shown on the top, middle and bottom panels, respectively. The index was selected by following criterion: reduced chi-square $< 1.2$ and degree of freedoms $> 20$. The grey shadows denote the main flares epochs, and the pink shadows mark the weak flares epochs.}\label{fig:x_result}
\end{figure}

The X-ray energy spectra were analyzed with the redshift power-law model (\texttt{zpowerlw}) through \texttt{xspec} version 12.10.1f. The mean value of the power-law index is $2.38\pm0.02$ with the maximum and minimum value of $2.98^{+0.47}_{-0.42}$ and $1.81^{+0.27}_{-0.24}$ respectively, and the results can be seen on the middle panel of Figure \ref{fig:x_result}. Centering at the peak time, we define the 80\% of the interval between the two predicted time of minimum luminosity as the flare epoch.
We can also define three states to describe the X-ray activity: quiescent, weak flares, and main flares, which corresponding to the different epochs in the results of the two-jet precession model.

To study the evolution of the X-ray flux of PG 1553+113, we analyzed the hardness ratio (HR) expressed as
\begin{equation}
HR = \frac{CR(2.0 \mbox{--} 10.0 \text{keV}) - CR(0.3 \mbox{--} 2.0 \text{keV})}{CR(2.0 \mbox{--} 10.0 \text{keV})+CR(0.3 \mbox{--} 2.0 \text{keV})},
\end{equation}
where CR(2.0--10.0 keV) and CR(0.3--2.0 keV) are counts rate of the hard and soft X-ray, respectively. The evolution of HR is shown on the bottom panel of Figure \ref{fig:x_result}. From 2012.0 to 2020.0, the average value of HR is $-0.689\pm 0.004$ with the maximum and minimum value are $-0.468 \pm 0.062$ and $-0.838\pm 0.058$ respectively, which exhibits a very soft level in the stage.

\begin{figure}
  \centering
  \includegraphics[angle=90,scale=0.3]{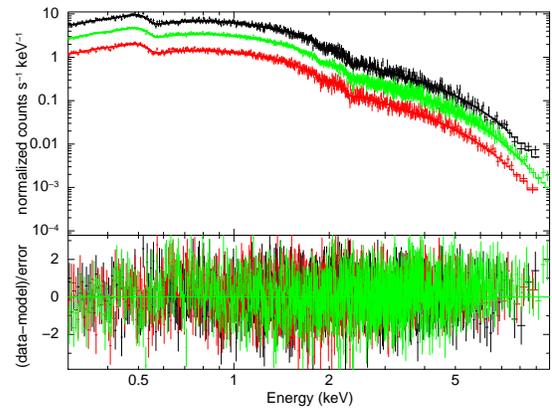}
  \caption{\emph{XMM-Newton} X-ray energy spectra for main flare (black), weak flare (green) and quiescent state (red). Three spectra were fitted by the \texttt{zpowerlw} model. The power-law indexes for the three states are $2.40\pm 0.01$ (weak flare), $2.44\pm0.02$ (main flare) and $2.52\pm0.01$ (quiescent state), respectively. The reduced chi-square for the spectral fitting are 1.00 (weak flare), 0.99 (main flare) and 1.00 (quiescent state), respectively. The errors are 90\% confidence.}\label{fig:xspec}
\end{figure}

We analyse X-ray data acquired by the \emph{XMM-Newton} Observatory.
The main flare spectrum was obtained in September 6, 2001 (ID 0094380801), the weak flare was observed in July 24, 2013 (ID 0727780101), and the spectrum for a quiescent state with observation ID 0810830201 was obtained in August 27, 2019. The spectra were analysed with \texttt{xspec}, and final results are presented in Figure \ref{fig:xspec}.
The X-ray energy spectra exhibit different behavior at these epochs.
\cite{raiteri2015} compared the spectral energy distribution (SEDs) for three states, and from their results, the quiescent state SEDs showed the significant different behavior from the flare states in $\gamma$-ray, which may be the contribution of the jet's activity.

\subsection{ Flare spectral indices}
\begin{figure*}
  \centering
  \subfigure[]{\includegraphics[width=0.4\textwidth]{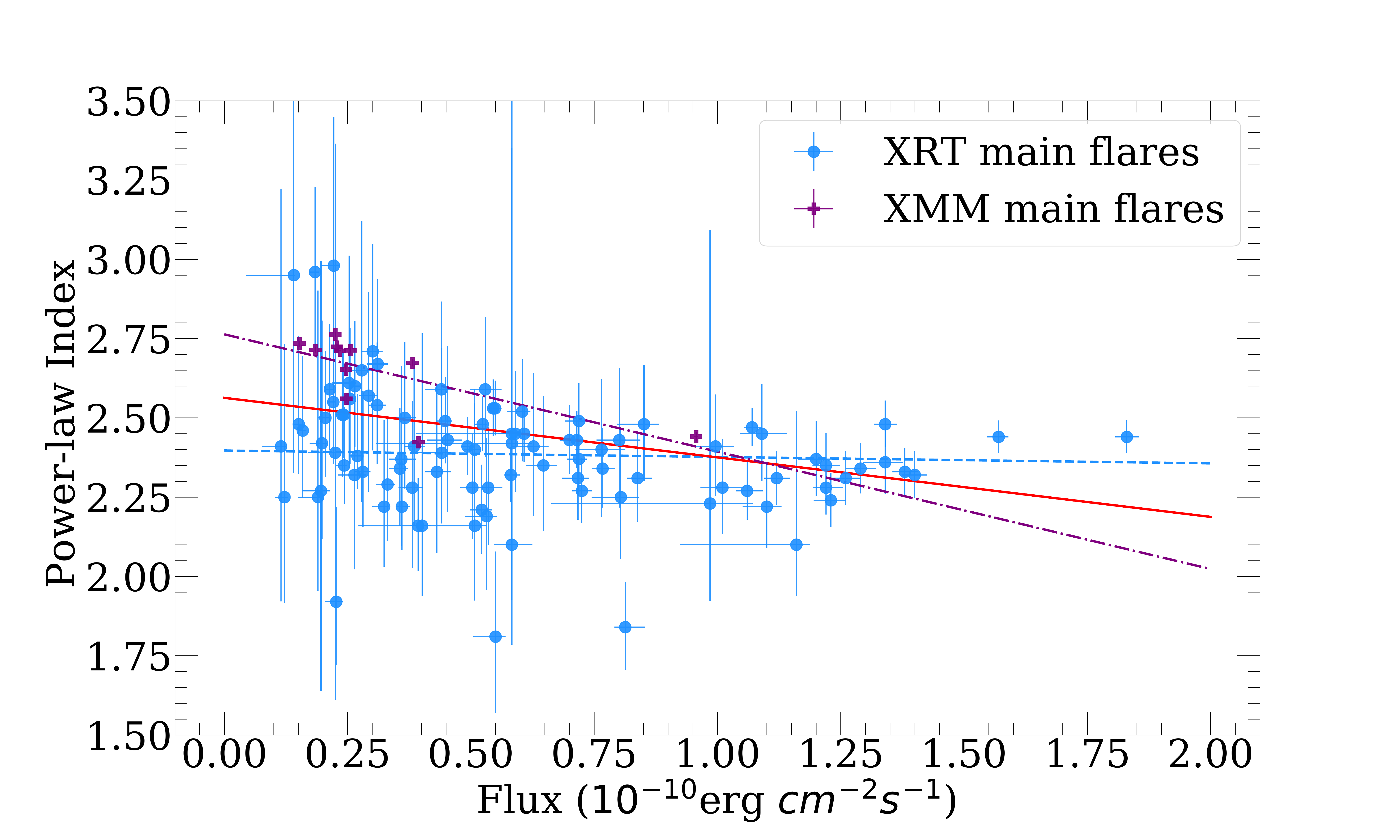}}
  \subfigure[]{\includegraphics[width=0.4\textwidth]{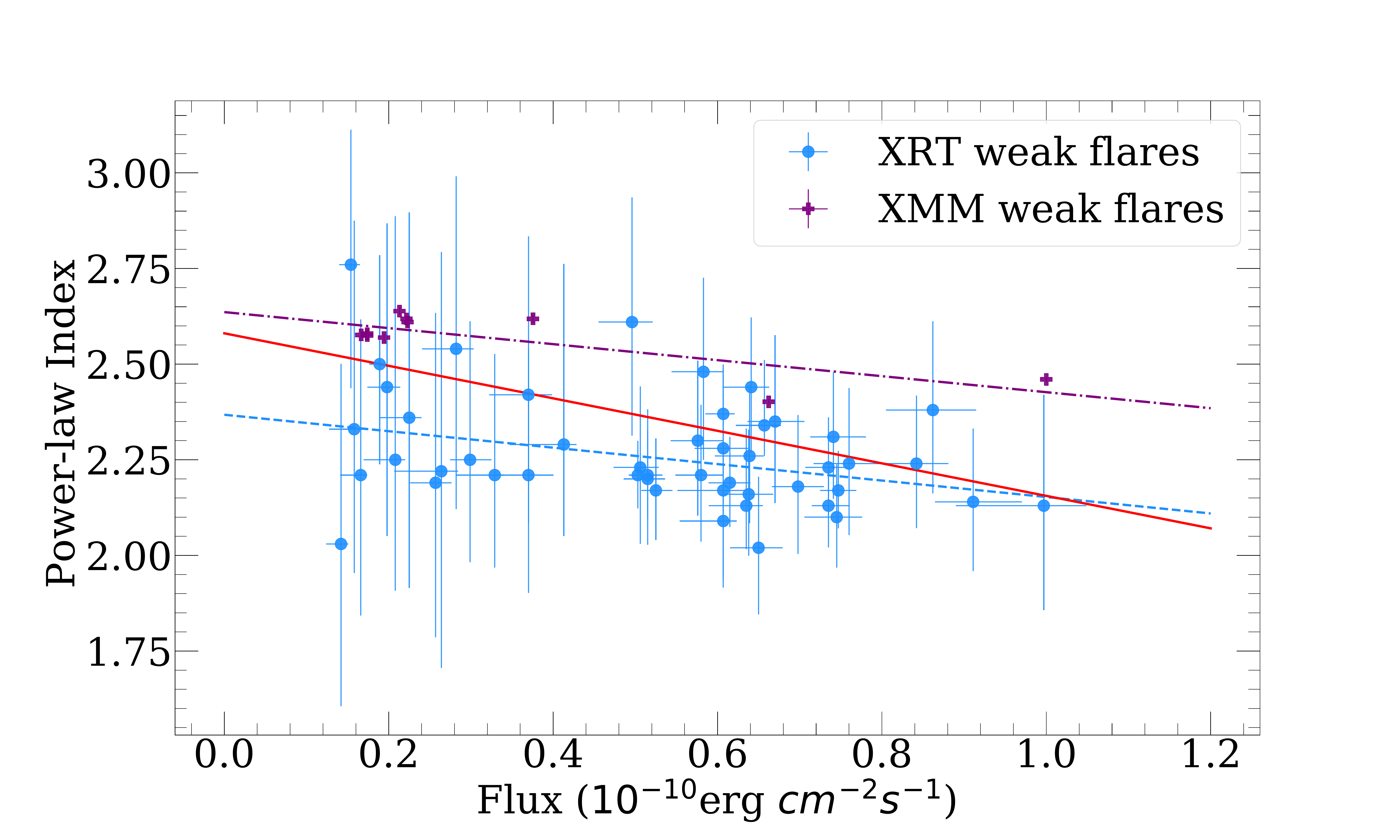}}
  \subfigure[]{\includegraphics[width=0.4\textwidth]{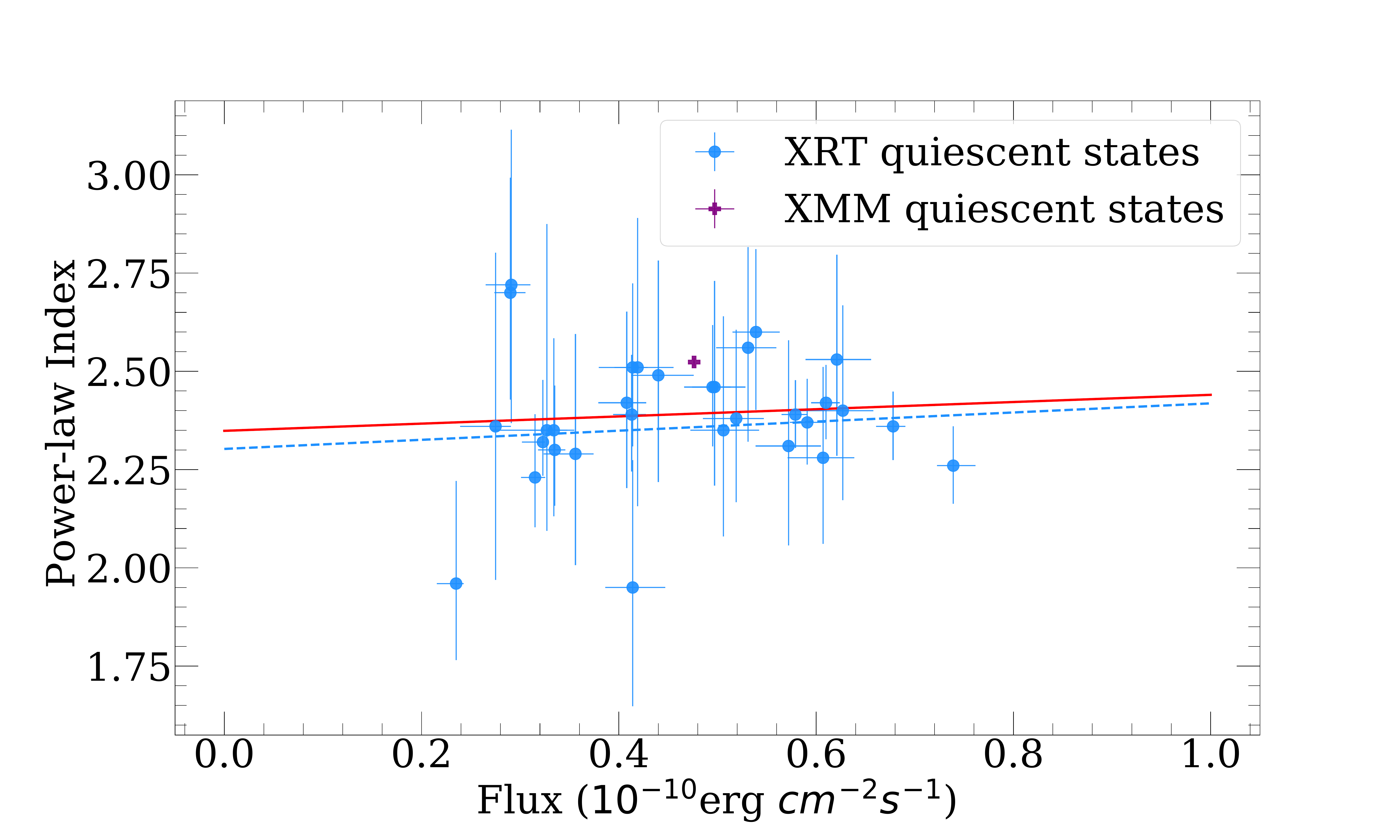}}
  \caption{The correlation between flux and power-law index for three different states. The results were obtained by spectral fitting with a single power-law, and then we selected the results whose reduced Chi-square $\chi_r^2 < 1.2$ and the degree of freedoms $>20$ are shown. To compare the discrepancy for these two telescopes, we fitted the data and presented the results from \emph{Swift/XRT} (blue) and \emph{XMM-Newton} (purple) separately. The red lines are the best fit of the flux and power-law index for the whole data, and the blue dash lines are the best fit of the \emph{Swift/XRT} data, the best fit for XMM data are shown as the purple dash-dot lines. }
  \label{fig:index_flux}
\end{figure*}

The power-law index and hardness ratio stay at a relatively stable level most of the time. The best linear fit of the correlation between flux and the power-law index are shown in Figure \ref{fig:index_flux}, and the results reveal a slightly ``harder-when-brighter" feature during the main and weak flares while for the quiescent state, the slightly ``softer-when-brighter" tendency can be seen. We explore the correlation between the index and flux with Spearman test. The analysis is processed by the python package \texttt{scipy.stats.spearmanr}. During main flares, the correlation coefficients ($r_s$) $< -0.2$ for all cases with p-value $< 0.05$. For the case of weak flares, $r_s < -0.1$ and the p-value $< 0.7$ (the result from \emph{XMM-Newton} shows the p-value of 0.675). The result of quiescent state reveals $r_s < 0.2$ with p-value $< 0.6$. It should be noted that in the quiescent state, we obtain only one data point from \emph{XMM-Newton}, hence the correlation test for this instrument can not be applied. More details for the linear fitting and Spearman test are shown on Table~\ref{tab:index_flux}.

When we fit all the data together, the slop is very different from the separated fitting results. We note that this behavior is related to the distribution of data points. 8 XMM data points for weak flares locate on the low flux and the power-law index $>2.5$. And for the case of XRT, few points locate on such interval. Therefore, the XMM data points decrease the slope when we fit the data from two instruments together.

The similar evolution of the spectral index for both main and weak flares suggest the same radiation mechanism, which may be related to the jets. The different behavior of the quiescent states and flares may be the contribution of the precession of jets. When the jet pointing to the observer, the ``harder-when-brighter" tendency in the flares states can be seen, and in the quiescent state, with the fraction of the contribution from the disc increase, the spectra tend to be ``softer-when-brighter" \citep{wang2020}.

\begin{deluxetable*}{cccccc}
\tablenum{2}
\tablecaption{The correlation between spectral indices and flux. \label{tab:index_flux}}
\tablewidth{0pt}
\tabletypesize{\scriptsize}
\tablehead{
 \colhead{Epoch} & \colhead{Instrument} & \colhead{$r_s$} & \colhead{P-Value}
& \colhead{Slope} & \colhead{Intercept} }
\decimalcolnumbers
\startdata
Main flares & XRT & -0.296 & 0.003 & $-0.020^{+0.006}_{-0.006}$ & $2.397^{+0.058}_{-0.057}$  \\
Main flares & XMM & -0.846 & 0.001 & $-0.370^{+0.269}_{-0.237}$ & $2.763^{+0.102}_{-0.103}$  \\
Main flares & All & -0.406 & $8.919\times 10^{-6}$ & $-0.187^{+0.081}_{-0.075}$ & $2.563^{+0.063}_{-0.067}$  \\ \hline
Weak flares & XRT & -0.308 & 0.033 & $-0.218^{+0.319}_{-0.320}$ & $2.369^{+0.202}_{-0.201}$  \\
Weak flares & XMM & -0.152 & 0.675 & $-0.210^{+0.137}_{-0.131}$ & $2.636^{+0.059}_{-0.060}$  \\
Weak flares & All & -0.415 & 0.001 & $-0.426^{+0.180}_{-0.178}$ & $2.582^{+0.102}_{-0.111}$  \\ \hline
Quiescent state & XRT & 0.107 & 0.567 & $0.115^{+0.382}_{-0.380}$ & $2.303^{+0.203}_{-0.204}$  \\
Quiescent state & XMM & - & - & - & -  \\
Quiescent state & All & 0.105 & 0.566 & $0.093^{+0.484}_{-0.466}$ & $2.349^{+0.238}_{-0.252}$  \\
\enddata
\tablecomments{The Spearman-rank test correlation coefficient, p-value, slope, intercept, respectively. The slope and intercept are obtained from the linear fitting. We compare the results of two instruments for each epoch. Then, the data of \emph{Swift} and \emph{XMM-Newton} are combined together in the analysis and the results are shown on the rows labeled with ``All".}
\end{deluxetable*}

\subsection{Some possible origins for X-ray light curves variation}
In the previous subsections, we argue that the X-ray variation is dominated by the precession of jets, however, the other potential contributions may be considered. We discuss the lighthouse effect, general relativity effect, disc instability, and Lense-Thirring precession below. But the timescale of the variation from these effects is inconsistent with the 2.2-year quasi-periodicity, however, we can not exclude out the modulation of these effects on the long and short timescale variability of PG 1553+113.

\subsubsection{The lighthouse effect}
With the magnetic field of the disc, the plasma bubbles may be injected into the jet and would be forced to move along it. As the result, the rotation of these bubbles around the jet would induce the variability of the light curves \citep{camenzind1992}. In this scenario, the variation of the light curves is determined by the viewing angle between the line of sight and the velocity direction of the bubbles and the angle should be time-dependent and is related to the spin of the black hole. The lighthouse effect can explain the intraday variability of 3C 273 and BL Lacertae well. However, the variability timescale of this effect is shorter compared to the quasi-periodicity of 2.2-year in PG 1553+113.

\subsubsection{General relativity effect on the jet}
When a blob is produced and propagate from disc to jet, its motion would be affected by the magnetic field, the radiation field, and the gravitational field which may induce the variation of the light curves \citep{abramowicz1990,mohan2015}. According to this scenario, the production of a blob is located in the vicinity of the black hole, as a result, some general relativity (GR) effects such as the gravitational redshift, gravitational lensing and the time delay should be taken into account. The helical orbit of the blobs cause the quasi-periodic variation of the light curves, and the results of \cite{mohan2015} suggest that its time scale ranges from few days to tens of days.

\subsubsection{Disc instability}
The whole SMBHB system may be surrounded by a huge disc so that the two black holes are embedded in it, and the periodic leakage of the gas into the cavity around the SMBHB may occur \citep{tanaka2013}. As a result, the fluctuation of flux which is produced by the periodic accretion can be observed. In this scenario, the variation time scale is related to the orbital period, however, the X-ray emission should be very soft and its light curve would decline as a power-law just similar to the case of tidal disruption events ($L \propto t^{-5/3}$). According to the fitting results, the emission of PG 1553+113 should be dominated by the jets, and the contribution of the disc is weaker at least by an order of magnitude. Hence we argue that the effect of the instability of the disc may be inconspicuous for PG 1553+113.

\subsubsection{Lense-Thirring precession}
The Lense-Thirring precession of the disc would induce the quasi-periodic oscillations (QPO) in the Kerr space time \citep{aschenbach2004,johannsen2011}. For the case of PG 1553+113, the time scale of QPO due to Lense-Thirring precession is much longer than the orbital period in SMBHB system \citep{sobacchi2017}. As a consequence, the effect from the Lense-Thirring precession only contributes to the long time scale period of this system.

\section{Conclusion}

Blazar PG 1553+113 is a SMBHB system candidate showing quasi-periodic variability across the electromagnetic spectrum. The physical origin of this variability is still not well-understood. We study the flux and spectral variability of PG 1553+113 on long-term timescales using \emph{Swift} and \emph{XMM-Newton} X-ray data collected for the period 2012--2020. During this period, the X-ray light curve of PG 1553+113 exhibits several main and weak flares. The main X-ray flares and some weak flares are consistent with the corresponding flares observed on the $\gamma$-ray band. A ``harder-when-brighter" behavior is observed in the X-ray for both main and weak flares, while a ``softer-when-brighter" behavior in quiescent states.

The observed quasi-periodic variation in the $\gamma$-ray light curves could be explained with one-jet model, however the origin of weak X-ray flares would remain unclear. In this study, we apply a two-jet model and show that the contribution from another jet and the accretion disc should be considered to explain the X-ray variability. Based on our fitting results for the X-ray light curve and the correlation between the disc and the jet \citep{ghisellini2014}, we obtain the mass of the primary black hole is $\simeq 3.47\times 10^8M_{\sun}$, the secondary one $\simeq 1.40\times 10^8M_{\sun}$, and the mass ratio $\simeq 0.41$ that is in line with previous studies. Although we can not obtain the exact values of the parameters, the order of magnitude of the total mass and the mass ratio are consistent with the previous works. In order to research the more detail of the variability, more high cadence observations are needed.

Blazar PG 1553+113 is not the first object with possible double-jet structure signatures. For example, \cite{qian2019} proposed a double-jet structure scenario for 3C 279 and suggested that there may be a SMBHB ejecting two precessing relativistic jets. At variance with their method, we focused on X-ray bands. This may be a indirect tool to search such SMBHB systems with two jets in the future.

In the extreme mass ratio black hole binary systems, the accretion disc may not form around the secondary black hole within the gravity of the primary one, such as OJ 287 \citep{lehto1996}. However, in mildly mass ratio SMBHB systems, it is possible that each black hole may have its own disc \citep{farris2014,bowen2017,ryan2017} and two-jet signature can be detected. Probably, our approach is valid in the case of the SMBHB systems with a mild mass ratio and we can detect such types of black hole systems in the future.

In complex environment of blazars, a pure geometric model is one of the tools to study the X-ray light curves. A simple model should not be expected to complete the explanation perfectly. Some intrinsic variation in the jets and discs should be considered in further research.

\acknowledgments

We are grateful to the anonymous referee for valuable comments that improved this work. This work is supported by the Natural Science Foundation of China under grant No.~11873035, the Natural Science Foundation of Shandong province (No.~JQ201702), and the Young Scholars Program of Shandong University (No.~20820162003).Yunguo Jiang is supported by Shandong Provincial Natural Science Foundation (No.~ZR2020MA062) and National Natural Science Foundation of China (No.~U2031102). S. Alexeeva acknowledges support from the National Natural Science Foundation of China (grant No.~12050410265) and LAMOST FELLOWSHIP program that is budgeted and administrated by Chinese Academy of Sciences. We acknowledge Swift for the observation and providing the public data for the research. OVRO 40-m monitoring program \citep{richards2011} which is supported in part by NASA grants NNX08AW31G, NNX11A043G, and NNX14AQ89G and NSF grants AST-0808050 and AST-1109911. Based on observations obtained with XMM-Newton, an ESA science mission with instruments and contributions directly funded by ESA Member States and NASA.

\software{astropy \citep{astropy2013,astropy2018}, emcee \citep{foreman2013}, matplotlib \citep{hunter2007}, scipy \citep{virtanen2020}}

\end{document}